\documentclass[11pt]{article}
\usepackage{typearea}\typearea{13}
\usepackage{amsmath,amsfonts,amssymb}
\usepackage[dvipdfmx]{graphicx}
\usepackage[dvipdfmx]{hyperref}\hypersetup{hidelinks}

\makeatletter
\long\def\@makecaption#1#2{{\small
\advance\leftskip1cm
\advance\rightskip1cm
\vskip\abovecaptionskip
\sbox\@tempboxa{#1: #2}%
\ifdim \wd\@tempboxa >\hsize
 #1: #2\par
\else
\global \@minipagefalse
\hb@xt@\hsize{\hfil\box\@tempboxa\hfil}%
\fi
\vskip\belowcaptionskip}}
\makeatother

\newcommand{\nl}{\notag\\}

\def\eq#1\en{\begin{equation}#1\end{equation}}  
\def\eqsplit#1\ensplit{
	\begin{equation}\begin{split}#1\end{split}\end{equation}
	}
\def\eqalign#1\enalign{
	\begin{align}#1\end{align}
	}
\def\eqa#1\ena{
	\begin{align}#1\end{align}
	}
\def\eqg#1\eng{
	\begin{gather}#1\end{gather}
}
\def\eqmul#1\enmul{
	\begin{multline}#1\end{multline}
	}
\newcommand{\lb}[1]  {\label{e:#1}}
\newcommand{\rlb}[1] {\eqref{e:#1}}     
\newtheorem{theorem}{Theorem}
\newtheorem{T}[theorem]{Theorem}

\newcommand{\snorm}[1]{\Vert#1\Vert}

\newcommand{\bbkt}[1]{\bigl\langle#1\bigr\rangle}

\newcommand{\sbkt}[1]{\langle#1\rangle}

\newcommand{\bra}[1]{\langle#1|}
\newcommand{\ket}[1]{|#1\rangle}

\newcommand{\sumtwo}[2]%
{\mathop{\sum_{#1}}_{#2}}
\newcommand{\sumthree}[3]%
{\mathop{\mathop{\sum_{#1}}_{#2}}_{#3}}
\newcommand{\sumfour}[4]%
{\mathop{\mathop{\mathop{\sum_{#1}}_{#2}}_{#3}}_{#4}} 
\newcommand{\prodtwo}[2]%
{\mathop{\prod_{#1}}_{#2}}
\newcommand{\mintwo}[2]%
{\mathop{\min_{#1}}_{#2}}
\newcommand{\maxtwo}[2]%
{\mathop{\max_{#1}}_{#2}}
\newcommand{\maxthree}[3]%
{\mathop{\mathop{\max_{#1}}_{#2}}_{#3}}
\newcommand{\limtwo}[2]%
{\mathop{\lim_{#1}}_{#2}}
\newcommand{\suptwo}[2]%
{\mathop{\sup_{#1}}_{#2}}
\newcommand{\supthree}[3]%
{\mathop{\mathop{\sup_{#1}}_{#2}}_{#3}}
\newcommand{\supfour}[4]%
{\mathop{\mathop{\mathop{\sup_{#1}}_{#2}}_{#3}}_{#4}} 
\newcommand{\inftwo}[2]%
{\mathop{\inf_{#1}}_{#2}}
\newcommand{\infthree}[3]%
{\mathop{\mathop{\inf_{#1}}_{#2}}_{#3}}
\newcommand{\inffour}[4]%
{\mathop{\mathop{\mathop{\inf_{#1}}_{#2}}_{#3}}_{#4}} 

\newcommand\calB{{\cal B}}

\newcommand\calH{{\cal H}}

\newcommand\calO{{\cal O}}

\newcommand{\ha}{\hat{a}}
\newcommand{\had}{\hat{a}^\dagger}

\newcommand{\hh}{\hat{h}}
\newcommand{\ho}{\hat{o}}

\newcommand{\hn}{\hat{n}}
\newcommand{\hp}{\hat{p}}

\newcommand{\hA}{\hat{A}}

\newcommand{\hH}{\hat{H}}

\newcommand{\hN}{\hat{N}}
\newcommand{\hP}{\hat{P}}

\newcommand{\bbR}{\mathbb{R}}
\newcommand{\bbZ}{\mathbb{Z}}
\newcommand{\ep}{\varepsilon}

\newcommand{\up}{\uparrow}
\newcommand{\dn}{\downarrow}

\newcommand{\pmat}[1]{\begin{pmatrix}#1\end{pmatrix}}

\newcommand{\sumLaL}{\sum_{x\in\LaL}}
\newcommand{\sumBL}{\sum_{(x,y)\in\BL}}

\newcommand{\hU}{\hat{U}}

\newcommand{\La}{\Lambda}

\newcommand{\LaL}{\Lambda}
\newcommand{\BL}{\calB}

\newcommand{\Zd}{\bbZ^d}

\newcommand{\hOL}{\hat{\calO}}
\newcommand{\hOLa}{\hat{\calO}^{(\alpha)}}
\newcommand{\hOLo}{\hat{\calO}^{(1)}}
\newcommand{\hOLt}{\hat{\calO}^{(2)}}

\newcommand{\hoo}{\ho^{(1)}}
\newcommand{\hot}{\ho^{(2)}}

\newcommand{\vac}{\ket{\Phi_\mathrm{vac}}}

\newcommand{\GS}{\ket{\Phi_\mathrm{GS}}}
\newcommand{\GSb}{\bra{\Phi_\mathrm{GS}}}

\newcommand{\EGS}{E_{\rm GS}}

\newcommand{\MmL}{M_{\rm max}(L)}

\newcommand{\ms}{m^*}

\newcommand{\limM}{\lim_{M\up\infty}}

\newcommand{\limL}{\lim_{L\up\infty}}

\newcommand{\Ga}{\ket{\Phi_{\rm GS}^{\rm a}}}
\newcommand{\Gb}{\ket{\Phi_{\rm GS}^{\rm b}}}

\newcommand{\fnorm}[1]{\frac{#1}{\snorm{#1}}}
\newcommand{\fnormt}[2]{\frac{#1#2}{\snorm{#2}}}

\newcommand{\hoa}{\ho_{\rm a}}
\newcommand{\hob}{\ho_{\rm b}}

\begin{document}

\noindent
{\Large\bf Spontaneous symmetry breaking in coupled Bose-Einstein \\condensates}
\par\bigskip

\renewcommand{\thefootnote}{\fnsymbol{footnote}}
\noindent
Hal Tasaki\footnote{
Department of Physics, Gakushuin University, Mejiro, Toshima-ku, 
Tokyo 171-8588, Japan
}
\renewcommand{\thefootnote}{\arabic{footnote}}\setcounter{footnote}{0}

\begin{quotation}
We study a system of two hardcore bosonic Hubbard models weakly coupled with each other by tunneling.
Assuming that the single uncoupled model exhibits off-diagonal long-range order, we prove that the coupled system exhibits  spontaneous symmetry breaking (SSB) in the infinite volume limit, in the sense that the two subsystems maintain a definite relative U(1) phase when the tunneling is turned off.
Although SSB of the U(1) phase is never observable in a single system, SSB of the relative U(1) phase is physically meaningful and observable by interference experiments.
The present theorem is made possible by the rigorous theory of low-lying states and SSB in quantum antiferromagnets developed over the years.
\end{quotation}

\section{Introduction}
The essence of Bose-Einstein condensation is off-diagonal long-range order (ODLRO) related to the global quantum mechanical U(1) phase \cite{Dalfovo,Leggett2001,BlochDalibardZwerger,PitaevskiiStringari}.
It is known however that spontaneous symmetry breaking (SSB) of the U(1) symmetry, which is predicted by mean filed-theories, is never observable because of the law of particle number conservation \cite{Leggett2001,LeggettSols}.
Although such a lack of SSB is sometimes regarded as paradoxical, there is indeed nothing problematic.
A ground state with long-range order (LRO) but without SSB is perfectly understood by the rigorous theory of low-lying states and SSB developed mainly for quantum antiferromagnets \cite{Anderson1952,Lhuillier,HorschLinden,KaplanHorschLinden,KomaTasaki1993,KomaTasaki1994,Tasaki2018,TasakiBOOK}.

In the present paper, we demonstrate, through a new rigorous result, that SSB of the relative U(1) symmetry may take place when two Bose-Einstein condensates are weakly coupled \cite{LeggettSols}.
More precisely we consider two hardcore bosonic Hubbard models coupled with each other through weak tunneling.
By making full use of the rigorous theory of low-lying states, we prove that, in the infinite volume limit, the two subsystems maintain definite relative U(1) phase when the tunneling is turned off, provided that the single hardcore bosonic Hubbard model exhibits ODLRO.
We stress that the SSB of relative U(1) phase is physically meaningful, and directly related to the observation of interference patterns in cold atom experiments \cite{Andrews,Gring}.
See \cite{Herring} for related phenomena in a classical setting, 

We have formulated our theorem for the hardcore bosonic Hubbard model mainly for simplicity.
It is in principle possible to extend the result to a broader class of boson systems, but that requires extra (nontrivial) work to extend the theorems in \cite{KomaTasaki1994,Tasaki2018}.

\section{Definition and ODLRO in the uncoupled system}
We start by defining the basic (uncoupled) hardcore bosonic Hubbard model, and state our assumption about ODLRO.
Let $\La$ be the $d$-dimensional $L\times\cdots\times L$ hypercubic lattice, and let $\BL$ be the corresponding set of  bonds $(x,y)$, i.e., ordered pairs of neighboring sites $x,y\in\La$.
We impose periodic boundary conditions.
For each site $x\in\LaL$ we denote by $\ha_x$ and $\had_x$ the annihilation and the creation operators, respectively, of a bosonic particle at site $x$.
They satisfy the standard commutation relations
$[\ha_x,\ha_y]=[\had_x,\had_y]=0$ and $[\ha_x,\had_y]=\delta_{x,y}$ for for any $x,y\in\LaL$.
The number operator at site $x$ is defined by $\hn_x=\had_x\ha_x$, and the total number operator by $\hN=\sumLaL\hn_x$.
We denote by $\vac$ the unique state such that $\ha_x\vac=0$ for any $x\in\LaL$.

Let $N=1,2,\ldots,L^d$ be the total boson number.
When we vary the system size, we always fix the density $\rho=N/L^d$, and make both $L$ and $N$ large.
We define the Hilbert space $\calH_N$ with $N$ hardcore bosons as the space spanned by all states  $\had_{x_1}\had_{x_2}\cdots\had_{x_N}\vac$ such that $x_i\ne x_j$ if $i\ne j$.
Then our Hamiltonian is
\eq
\hH=-\hP_{\rm hc}\sumBL\had_x\ha_y,
\lb{BEC3B}
\en 
where $\hP_{\rm hc}$ is the projection operator onto $\calH_N$.
We denote by $\GS\in\calH_N$ the unique ground state of $\hH$.
The uniqueness is proved by the standard Perron-Frobenius argument \cite{Marshall,LiebMattis1962}.
See \cite{TasakiBOOK}.

To test for possible ODLRO, we define the order operators by
\eq
\hOL^+=\sumLaL\had_x,\quad\hOL^-=\sumLaL\ha_x,
\lb{BEC7}
\en
and also the self-adjoint order operators by
\eq
\hOLo=\frac{\hOL^++\hOL^-}{2},\quad\hOLt=\frac{\hOL^+-\hOL^-}{2i}.
\lb{BEC9}
\en
It is readily found that these operators are transformed by the global U(1) phase rotation $\hU_\theta=e^{-i\theta\hN}$ as 
\eq
\hU_\theta^\dagger\hOLo\hU_\theta=\cos\theta\,\hOLo-\sin\theta\,\hOLt,\quad
\hU_\theta^\dagger\hOLt\hU_\theta=\cos\theta\,\hOLt+\sin\theta\,\hOLo.
\lb{BEC10}
\en
This means that the pair $(\hOLo,\hOLt)$ transforms precisely as a vector.
We shall assume throughout the present paper that, for some dimension $d$ and the density $\rho\in(0,1)$, the ground state $\GS$ of the hard-core bosonic Hubbard model \rlb{BEC3B} exhibits ODLRO in the sense that
\eq
\GSb\Bigl(\frac{\hOLo}{L^d}\Bigr)^2\GS=\GSb\Bigl(\frac{\hOLt}{L^d}\Bigr)^2\GS\ge \lambda(\rho),
\lb{BEC11}
\en
for any $L$ with the order parameter $\lambda(\rho)>0$.\footnote{%
We here take the ``statistical mechanical point of view'', and regard the Bose-Einstein condensation as a phenomenon in the infinite volume limit.}
This is expected to be valid for a large range of $\rho$ for any $d\ge2$, but is proved rigorously only for $d\ge2$ and $\rho=1/2$ by Kennedy, Lieb, and Shastry \cite{KennedyLiebShastry1988a,KennedyLiebShastry1988b}, and by Kubo and Kishi \cite{KuboKishi1988}, who extended the reflection positivity method due to Dyson, Lieb, and Simon \cite{DysonLiebSimon1978} and Neves and Perez \cite{NevesPerez1986}.
We note, on the other hand, that $\GS\in\calH_N$ implies
\eq
\GSb\frac{\hOLo}{L^d}\GS=\GSb\frac{\hOLt}{L^d}\GS=0,
\lb{BEC19}
\en
for any $d$, $\rho$, and $L$.

The ground state $\GS$ therefore exhibits (OD)LRO as in \rlb{BEC11} but no SSB as in \rlb{BEC19}.
Such LRO without SSB has been studied intensively in the context of quantum antiferromagnets, and it is known that, in such a situation, there inevitably appears a series of low-lying states, i.e., states with very low excitation energies, and physical ground states with SSB are linear combinations of the low-lying states \cite{Anderson1952}.
See, e.g., \cite{Lhuillier}.
For a discussion of similar phenomena in nuclear physics, see Chapter~11 of \cite{RingSchuck}.
By now a fully rigorous theory of low-lying states has been developed \cite{HorschLinden,KaplanHorschLinden,KomaTasaki1993,KomaTasaki1994,Tasaki2018,TasakiBOOK}.
We recommend \cite{Tasaki2018}, which can be read as a compact review.

\section{Symmetry breaking ground states in the uncoupled system}
\label{ss:BEC3}
We now review the theory of low-lying states and symmetry breaking in the context of hardcore bosonic Hubbard model.
To develop a theory parallel to that for quantum antiferromagnets, we introduce the extended  Hilbert space
\eq
\calH=\bigoplus_{K=0}^{L^d}\calH_{K},
\lb{BEC20}
\en
which contains all possible particle numbers.
We then take the Hamiltonian
\eq
\hH_\mu=\hH-\mu\hN,
\lb{BEC21}
\en
on $\calH$, and adjust the chemical potential $\mu$ so that the ground state of $\hH_\mu$ coincides with the previous ground state $\GS$ which has the fixed density $\rho$.
We assume that such tuning of $\mu$ is possible.
When the target density is $\rho=1/2$, it is known rigorously that the right choice is $\mu=0$ \cite{TasakiBOOK,Marshall,LiebMattis1962,Mattis79,Nishimori81}.

Following \cite{KomaTasaki1994,Tasaki2018}, we define trial states 
\eq
\ket{\Gamma_M}=\frac{(\hOL^+)^M\GS}{\snorm{(\hOL^+)^M\GS}},\quad
\ket{\Gamma_{-M}}=\frac{(\hOL^-)^M\GS}{\snorm{(\hOL^-)^M\GS}},
\lb{BEC22}
\en
for $M=1,2,\ldots$.
The following theorem, which establishes the existence of the series of low-lying states, was proved in \cite{KomaTasaki1994} for the case $\rho=1/2$, and in \cite{Tasaki2018} for general $\rho$.
\begin{T}
\label{t:KomaTasaki4}
Suppose that the ground state $\GS$ exhibits ODLRO as in  \rlb{BEC11}.
Then there are constants $C_1$ and $C_2$ which depend only on $d$, $\rho$, and $\lambda(\rho)$.
For any $L$ and $M$ such that $|M|\le C_1L^{d/2}$, the state $\ket{\Gamma_M}$ is well-defined, and satisfies\footnote{%
If $\rho=1/2$, the second term in the right-hand side can be replaced by $C_2M^2/L^d$.
We expect that the same bound is possible for general $\rho$, but cannot prove it for technical reasons.
}
\eq
\bra{\Gamma_M}\hH_\mu\ket{\Gamma_M}\le\GSb\hH_\mu\GS+C_2\frac{|M|^3}{L^d}.
\lb{BEC23}
\en
\end{T}

We proceed  to construct a new trial state by summing up these low-lying states.
For $\theta\in\bbR$ and  an integer valued function $\MmL>0$ such that $\MmL\le C_1L^{d/2}$, let
\eq
\ket{\Xi_\theta}=\frac{1}{\sqrt{2\MmL+1}}\sum_{n=-\MmL}^{\MmL}
e^{-in\theta}\ket{\Gamma_n},
\lb{BEC24}
\en
where we set $\ket{\Gamma_0}=\GS$.
Let $x,y\in\bbZ^d$ be neighboring sites, and let $\hh_{x,y}=-\hP_{\rm hc}(\had_x\ha_y+\had_x\ha_y)$ be the local Hamiltonian.
By using \rlb{BEC23} and the translation invariance, one finds that
\eq
\limL\bra{\Xi_\theta}\hh_{x,y}\ket{\Xi_\theta}=\epsilon_{\rm GS}(\mu),
\lb{GSEd}
\en
where $\epsilon_{\rm GS}(\mu)=\limL\GSb\hH_\mu\GS/(dL^d)$ is the ground state energy per bond.
Thus the state $\ket{\Xi_\theta}$ can be regarded essentially as a ground state when $L$ is large.
Then the following theorem was proved in \cite{Tasaki2018}, improving the results in \cite{KomaTasaki1994}.

\begin{T}
\label{t:KomaTasaki5}
If $\MmL$ diverges to infinity not too rapidly as $L\up\infty$, one has
\eqg
\limL\bra{\Xi_\theta}\frac{\hOL^\pm}{L^d}\ket{\Xi_\theta}=\ms\,e^{\pm i\theta},
\lb{BEC25}\\
\limL\bra{\Xi_\theta}\frac{\hOLa}{L^d}\ket{\Xi_\theta}=
\begin{cases}
\ms\cos\theta&(\alpha=1)\\
\ms\sin\theta&(\alpha=2),\\
\end{cases}
\lb{BEC26}\\
\limL\bra{\Xi_\theta}\Bigl(\frac{\hOLa}{L^d}\Bigr)^2\ket{\Xi_\theta}=
\begin{cases}
(\ms\cos\theta)^2&(\alpha=1)\\
(\ms\sin\theta)^2&(\alpha=2),\\
\end{cases}
\lb{BEC27}
\eng
where the symmetry breaking order parameter $\ms$ is defined by
\eq
\ms:=\lim_{k\up\infty}\limL\Bigl\{
\GSb\Bigl(\frac{\hOLa}{L^d}\Bigr)^{2k}\GS
\Bigr\}^{1/(2k)},
\lb{LSHA6}
\en
with $\alpha=1,2$, and satisfies $\ms\ge\sqrt{2\lambda(\rho)}$.
\end{T}

The theorem shows that the state $\ket{\Xi_\theta}$, which is essentially a ground state of  \rlb{BEC21}, exhibits ODLRO and also fully breaks the U(1) phase symmetry.
The symmetry breaking is manifest in the remarkable relations
\eq
\bra{\Xi_\theta}\had_x\ket{\Xi_\theta}\simeq\ms\,e^{i\theta},\quad
\bra{\Xi_\theta}\ha_x\ket{\Xi_\theta}\simeq\ms\,e^{-i\theta},
\lb{BEC28}
\en
which hold for large $L$ because of \rlb{BEC25}.
In the state $\ket{\Xi_\theta}$, the U(1) phase is ``pointing'' in the specific direction $\theta$.

Comparing \rlb{BEC26} and \rlb{BEC27}, we find that  $\hOLa/L^d$, which is the density of the order operator, exhibits vanishing fluctuation as $L$ becomes large.
Vanishing fluctuation is usually a sign that the state is a physically realistic macroscopic state.
In fact the state corresponding to $\ket{\Xi_\theta}$ is regarded as a realizable ground state in quantum antiferromagnets.
Moreover the (near) ground state  $\ket{\Xi_\theta}$ is similar in many aspects to ground states obtained by mean-field theories for bosons \cite{Dalfovo,Leggett2001,PitaevskiiStringari,FetterWalecka}.
Nevertheless we cannot regard  $\ket{\Xi_\theta}$ as a ``realistic'' state of an isolated Bose-Einstein condensate since it is a superposition of states with different particle numbers.
(See, e.g., section~III.D.1 of \cite{Leggett2001} and \cite{LeggettSols}.)
Suppose that one confines exactly $N$ particles in a container and then cool them down to the ground state.
It is never possible to generate a superposition as in \rlb{BEC24} by allowed physical processes.
But a more important point is that the phase $\theta$ characterizing $\ket{\Xi_\theta}$ is a physically meaningless quantity, which can never be measured experimentally.

In this sense the exact ground state $\GS$, which exhibits ODLRO but no symmetry breaking, may be a better model of an isolated Bose-Einstein condensate.
We should note however that the distinction between $\ket{\Xi_\theta}$ and $\GS$ becomes subtle when we restrict the class of observables.
More precisely, if $\hA$ is a local observable such that $[\hN,\hA]=0$, we expect that 
\eq
\GSb\hA\GS\simeq\bra{\Xi_\theta}\hA\ket{\Xi_\theta}.
\lb{BEC37}
\en
We also note that the $\GS$  is obtained from  $\ket{\Xi_\theta}$ by
\eq
\GS=\sqrt{2\MmL+1}\,\hP_N\,\ket{\Xi_\theta},
\lb{BEC29A}
\en
where $\hP_N$ is the projection onto $\calH_N$, or by
\eq
\GS=\frac{\sqrt{2\MmL+1}}{2\pi}\int_0^{2\pi}d\theta\,\ket{\Xi_\theta}.
\lb{BEC29}
\en
It is interesting that the exact ground state $\GS$ can be regarded as  a superposition of the symmetry breaking (near) ground states  $\ket{\Xi_\theta}$ for all possible $\theta$.
See \cite{TasakiBOOK} for further discussion about the relation between $\GS$ and $\ket{\Xi_\theta}$.

\section{Coupled Bose-Einstein condensates}
\label{ss:BEC5}
We now consider a system of two Bose-Einstein condensates coupled weakly by tunneling.
We then find that a spontaneous symmetry breaking that fixes the relative U(1) phase takes place.
Unlike the phase $\theta$ discussed above, the relative phase $\varphi$ is experimentally measurable.
This situation corresponds, e.g., to the experimental setup where bosons are trapped in a double-well potential \cite{Andrews,Gring}.
The clear interference pattern observed experimentally \cite{Andrews} is a manifestation of a fixed relative U(1) phase.\footnote{%
The interference pattern is observed after switching off the trapping potential and letting the  particles evolve almost freely.
We should note that there is an essentially different class of interference phenomena between two Bose-Einstein condensates.
It is known that two condensates which have no fixed relative phase (and hence  well approximated by $\GS\otimes\GS$) also exhibit interference.
See, e.g., \cite{Javanainen}.
}

We consider two exact copies of the $d$-dimensional hyper cubic lattice $\LaL$, and call them $\La_{\rm a}$ and $\La_{\rm b}$.
Lattice sites are denoted as $(x,\nu)\in\La_\nu$ where $\nu=\rm a,b$ and $x\in\LaL$.
On each lattice we define the same system of hardcore bosons as before.
We further assume that there is a tunneling Hamiltonian which weakly couples the two subsystems on $\La_{\rm a}$ and $\La_{\rm b}$.
The total Hamiltonian is thus
\eq
\hH^{\rm tot}_\ep=\hH_{\rm a}+\hH_{\rm b}+\ep\hH_{\rm tunnel}.
\lb{BEC40}
\en
Here we set, for $\nu={\rm a,b}$,
\eq
\hH_\nu=-\hP_{\rm hc}\sumBL\had_{(x,\nu)}\ha_{(y,\nu)},
\lb{BEC41}
\en
which are the exact copies of \rlb{BEC3B}, and 
\eq
\hH_{\rm tunnel}=-\sumLaL\bigl(e^{i\varphi}\,\had_{(x,{\rm a})}\ha_{(x,{\rm b})}+e^{-i\varphi}\,\ha_{(x,{\rm a})}\had_{(x,{\rm b})}\bigr).
\lb{BEC42}
\en
Here the phase factor $\varphi\in\bbR$ is introduced to make clear the physical picture; it is most natural to set $\varphi=0$.\footnote{%
The general case reduces to $\varphi=0$ by replacement $e^{i\varphi}\ha_{(x,{\rm b})}\to\ha_{(x,{\rm b})}$ for all $x\in\La$.
}

We treat this problem in a physically realistic Hilbert space where the total number of particles in the coupled system is exactly $2N$.
If we denote the copy of the $N$ particle Hilbert space $\calH_N$ as $\calH_N^{\nu}$ for $\nu={\rm a,b}$, the whole Hilbert space is
\eq
\calH_{2N}^{\rm tot}=\bigoplus_{K=0}^{2N}\calH_K^{{\rm a}}\otimes\calH_{2N-K}^{{\rm b}}.
\lb{BEC43}
\en
In other words, we assume that the two Bose-Einstein condensates can exchange particles in a coherent manner, while completely isolated from the outside world.
This may be a reasonable idealization of realistic situations in cold atom experiments.

Let $\ket{\Phi^{\rm tot}_{{\rm GS},\ep}}\in\calH_{2N}^{\rm tot}$ be the unique ground state of the total Hamiltonian \rlb{BEC40} with $\varepsilon>0$, where the uniqueness is again easily proved by the Perron-Frobenius method. 
(See, e.g., \cite{TasakiBOOK}.) 
Then the main result of the present paper is the following.
\begin{T}
Assume the existence of ODLRO (in the single uncoupled system) as in \rlb{BEC11}.
Then for any $x\in\Zd$, we have
\eqg
\lim_{\ep\dn0}\limL\bra{\Phi^{\rm tot}_{{\rm GS},\ep}}\had_{(x,{\rm a})}\ha_{(x,{\rm b})}\ket{\Phi^{\rm tot}_{{\rm GS},\ep}}=\tilde{m}^2\,e^{-i\varphi},\\
\lim_{\ep\dn0}\limL\bra{\Phi^{\rm tot}_{{\rm GS},\ep}}\ha_{(x,{\rm a})}\had_{(x,{\rm b})}\ket{\Phi^{\rm tot}_{{\rm GS},\ep}}=\tilde{m}^2\,e^{i\varphi},
\lb{BEC44}
\eng
with $\tilde{m}\ge\ms\ge\sqrt{2\lambda(\rho)}$.
\end{T}
As we noted before the assumption of the theorem is rigorously established for $d\ge2$ and $N=L^d/2$.

The relation \rlb{BEC44} indicates that the two condensates are coupled in a coherent manner (or entangled) to have a definite relative U(1) phase.
To see this, define, for $\nu={\rm a,b}$ and $x\in\LaL$, the local order operators $\hoo_{(x,\nu)}$ and $\hot_{(x,\nu)}$ by
\eq
\hoo_{(x,\nu)}:=\frac{\had_{(x,\nu)}+\ha_{(x,\nu)}}{2},\quad
\hot_{(x,\nu)}:=\frac{\had_{(x,\nu)}-\ha_{(x,\nu)}}{2i}.
\lb{BEC45}
\en
Exactly as in \rlb{BEC9} and \rlb{BEC10}, the pair $(\hoo_{(x,\nu)},\hot_{(x,\nu)})$ transforms as a vector under the operation of the unitary operator $\hU_\theta=e^{-i\theta\hN}$.
We of course have $\bra{\Phi^{\rm tot}_{{\rm GS},\ep}}(\hoo_{(x,\nu)},\hot_{(x,\nu)})\ket{\Phi^{\rm tot}_{{\rm GS},\ep}}=(0,0)$ again by particle number conservation, but \rlb{BEC44} implies that 
\eq
\lim_{\ep\dn0}\limL\bra{\Phi^{\rm tot}_{{\rm GS},\ep}}
\bigl\{\hoo_{(x,{\rm a})}\hoo_{(x,{\rm b})}+\hot_{(x,{\rm a})}\hot_{(x,{\rm b})}\bigr\}
\ket{\Phi^{\rm tot}_{{\rm GS},\ep}}=\tilde{m}^2\,\cos\varphi.
\lb{BEC46}
\en
This suggests that the two order operators behaves like two vectors with magnitude $\tilde{m}$ which have a fixed relative angle $\varphi$.
This is more directly seen by noting that
\eq
\lim_{\ep\dn0}\limL\bra{\Phi^{\rm tot}_{{\rm GS},\ep}}(\hoo_{(x,{\rm a})},\hot_{(x,{\rm a})})
\pmat{\cos\varphi&\sin\varphi\\-\sin\varphi&\cos\varphi}
\pmat{\hoo_{(x,{\rm b})}\\\hot_{(x,{\rm b})}}
\ket{\Phi^{\rm tot}_{{\rm GS},\ep}}=\tilde{m}^2,
\lb{BEC47}
\en
which also follows from \rlb{BEC44}.

We conclude that, in the ground state obtained in the double limit $\lim_{\ep\dn0}\limL$, the relative U(1) phase between the two condensates has a definite value $\varphi$.
Since this phase ordering was achieved by ``infinitesimal symmetry breaking field'' $\ep$ in the tunneling Hamiltonian, we can say that this is a kind of spontaneous symmetry breaking.\footnote{%
Recall that in the ferromagnetic Ising model, for example, one considers a similar double limit $\lim_{h\dn0}\limL\sbkt{\sigma_x}$ to detect possible spontaneous symmetry breaking, where $h$ is the external magnetic field.
}
As we noted in the beginning, such ordering of relative phase between two weakly coupled Bose-Einstein condensates can be experimentally observed by means of interference experiments.\footnote{%
Although our theorem is about the $L\up\infty$ limit, it suggests that one should start seeing the phase ordering for $\ep\gtrsim L^{-2d}$ when $L$ is large but finite.
}

\section{Proof}
Note that the Hamiltonian \rlb{BEC40} is invariant under the transformation
\eq
\ha_{(x,\rm a)}\to e^{i\varphi}\ha_{(x,\rm b)},\quad\ha_{(x,\rm b)}\to e^{-i\varphi}\ha_{(x,\rm a)},
\lb{ab}
\en
for all $x\in\La$, and so is the unique ground state $\ket{\Phi^{\rm tot}_{{\rm GS},\ep}}$.
The invariance implies that 
\par\noindent$\bra{\Phi^{\rm tot}_{{\rm GS},\ep}}e^{i\varphi}\had_{(x,{\rm a})}\ha_{(x,{\rm b})}\ket{\Phi^{\rm tot}_{{\rm GS},\ep}}=\bra{\Phi^{\rm tot}_{{\rm GS},\ep}}e^{-i\varphi}\ha_{(x,{\rm a})}\had_{(x,{\rm b})}\ket{\Phi^{\rm tot}_{{\rm GS},\ep}}$, and hence this quantity is real.

The key of the proof is that we can take a state $\ket{\Theta^{L,M}_\varphi}\in\calH_{2N}^{\rm tot}$ for $\varphi\in\bbR$ with the following properties.
Like the ground state, it satisfies \par\noindent$\bra{\Theta^{L,M}_\varphi}e^{i\varphi}\had_{(x,{\rm a})}\ha_{(x,{\rm b})}\ket{\Theta^{L,M}_\varphi}=\bra{\Theta^{L,M}_\varphi}e^{-i\varphi}\ha_{(x,{\rm a})}\had_{(x,{\rm b})}\ket{\Theta^{L,M}_\varphi}\in\bbR$ and this quantity is independent of $x$.
It further satisfies 
\eq
\limM\limL\bra{\Theta^{L,M}_\varphi}e^{i\varphi}\had_{(x,{\rm a})}\ha_{(x,{\rm b})}\ket{\Theta^{L,M}_\varphi}=(\ms)^2
\lb{ms2}
\en
for any $x$.
Finally $\ket{\Theta^{L,M}_\varphi}$ is essentially a ground state of $\hH^{\rm tot}_0=\hH_{\rm a}+\hH_{\rm b}$ in the sense that 
\eq
\limL\frac{1}{L^d}\bigl\{\bra{\Theta^{L,M}_\varphi}\hH^{\rm tot}_0\ket{\Theta^{L,M}_\varphi}-E^{\rm tot}_{{\rm GS},0}\bigr\}=0,
\lb{LLS}
\en
for any fixed $M$, where $E^{\rm tot}_{{\rm GS},0}$ is the ground state energy of $\hH^{\rm tot}_0$ in the space $\calH_{2N}^{\rm tot}$.

Let us first assume the existence of $\ket{\Theta^{L,M}_\varphi}$, and prove the theorem by following the variational argument due to Kaplan, Horsch, and von der Linden \cite{KaplanHorschLinden}.
Since $\ket{\Phi^{\rm tot}_{{\rm GS},\ep}}$ is the ground state, one obviously has
\eq
\bra{\Theta^{L,M}_\varphi}\hH^{\rm tot}_\ep\ket{\Theta^{L,M}_\varphi}
\ge
\bra{\Phi^{\rm tot}_{{\rm GS},\ep}}\hH^{\rm tot}_\ep\ket{\Phi^{\rm tot}_{{\rm GS},\ep}}.
\en
Since $\hH^{\rm tot}_\ep=\hH^{\rm tot}_0+\ep\hH_{\rm tunnel}$, we have
\eqa
-\frac{1}{L^d}&\bra{\Phi^{\rm tot}_{{\rm GS},\ep}}\hH_{\rm tunnel}\ket{\Phi^{\rm tot}_{{\rm GS},\ep}}
\nl&\ge
-\frac{1}{L^d}\bra{\Theta^{L,M}_\varphi}\hH_{\rm tunnel}\ket{\Theta^{L,M}_\varphi}
+\frac{1}{\ep L^d}\bigl\{
\bra{\Phi^{\rm tot}_{{\rm GS},\ep}}\hH^{\rm tot}_0\ket{\Phi^{\rm tot}_{{\rm GS},\ep}}
-
\bra{\Theta^{L,M}_\varphi}\hH^{\rm tot}_0\ket{\Theta^{L,M}_\varphi}
\bigr\}
\nl&\ge
-\frac{1}{L^d}\bra{\Theta^{L,M}_\varphi}\hH_{\rm tunnel}\ket{\Theta^{L,M}_\varphi}
+\frac{1}{\ep L^d}\bigl\{
E^{\rm tot}_{{\rm GS},0}
-
\bra{\Theta^{L,M}_\varphi}\hH^{\rm tot}_0\ket{\Theta^{L,M}_\varphi}
\bigr\}.
\ena
By recalling \rlb{BEC42} and noting the symmetry, this becomes
\eq
\bra{\Phi^{\rm tot}_{{\rm GS},\ep}}e^{i\varphi}\had_{(x,{\rm a})}\ha_{(x,{\rm b})}\ket{\Phi^{\rm tot}_{{\rm GS},\ep}}
\ge
\bra{\Theta^{L,M}_\varphi}e^{i\varphi}\had_{(x,{\rm a})}\ha_{(x,{\rm b})}\ket{\Theta^{L,M}_\varphi}
+\frac{1}{2\ep L^d}\bigl\{
E^{\rm tot}_{{\rm GS},0}
-
\bra{\Theta^{L,M}_\varphi}\hH^{\rm tot}_0\ket{\Theta^{L,M}_\varphi}
\bigr\}.
\en
By fixing arbitrary $\ep>0$, letting $L\up\infty$, and then letting $M\up\infty$,  we get
\eq
\limL\bra{\Phi^{\rm tot}_{{\rm GS},\ep}}e^{i\varphi}\had_{(x,{\rm a})}\ha_{(x,{\rm b})}\ket{\Phi^{\rm tot}_{{\rm GS},\ep}}\ge(\ms)^2,
\en
where we used \rlb{LLS} and then \rlb{ms2}.
This becomes the desired \rlb{BEC44} if we let $\ep\dn0$.

\bigskip

We now move onto our main task of constructing $\ket{\Theta^{L,M}_\varphi}$.
Let $\ket{\Xi_\theta^{\rm a}}$ and $\ket{\Xi_\theta^{\rm b}}$ be the exact copies of $\ket{\Xi_\theta}$ defined in \rlb{BEC24}, which is essentially a ground state and also breaks U(1) symmetry.
A natural candidate for a state with fixed relative phase is the tensor product $\ket{\Xi_\theta^{\rm a}}\otimes\ket{\Xi_{\theta+\varphi}^{\rm b}}$, but this state is again a superposition of states with different total particle numbers.
We can follow \rlb{BEC29A} or \rlb{BEC29} to construct a physical state as 
\eq
\ket{\Xi^{\rm tot}_\varphi}\propto\hP^{\rm tot}_{2N}\bigl(\ket{\Xi_\theta^{\rm a}}\otimes\ket{\Xi_{\theta+\varphi}^{\rm b}}\bigr),
\lb{BEC48A}
\en
where $\hP^{\rm tot}_{2N}$ is the projection onto $\calH_{2N}^{\rm tot}$, or by phase averaging as
\eq
\ket{\Xi_\varphi^{\rm tot}}\propto\frac{1}{2\pi}\int_0^{2\pi}d\theta\,\ket{\Xi_\theta^{\rm a}}\otimes\ket{\Xi_{\theta+\varphi}^{\rm b}}.
\lb{BEC48}
\en
The two constructions lead to exactly the same result, and we get 
\eq
\ket{\Xi_\varphi^{\rm tot}}=\frac{1}{\sqrt{2\MmL+1}}\sum_{n=-\MmL}^{\MmL}
e^{in\varphi}\ket{\Gamma^{\rm a}_n}\otimes\ket{\Gamma^{\rm b}_{-n}},
\lb{BEC49}
\en
where $\ket{\Gamma^{\rm a}_M}$ and $\ket{\Gamma^{\rm b}_M}$ are exact copies of  low-lying states \rlb{BEC22}.
This is indeed a near ground state of $\hH^{\rm tot}_0$ in which the two condensates have definite relative phase $\varphi$.
For our purpose it is convenient to replace $\MmL$ by $M$ to define
\eq
\ket{\Theta_\varphi^{L,M}}=\frac{1}{\sqrt{2M+1}}\sum_{n=-M}^{M}
e^{in\varphi}\ket{\Gamma^{\rm a}_n}\otimes\ket{\Gamma^{\rm b}_{-n}}.
\lb{Theta}
\en
The desired properties \rlb{ms2} and \rlb{LLS} are proved by using Theorems~1 and 2.
See Appendices~\ref{a:ms} and \ref{a:en}

\section{Discussion}
We considered two systems of hardcore bosons weakly coupled with each other by the tunneling Hamiltonian \rlb{BEC42}.
Under the assumption that the uncoupled system exhibits ODLRO, we proved that, in the infinite volume limit, the two Bose-Einstein condensates maintain definite relative phase $\varphi$ when the tunneling is turned off.
This is naturally interpreted as spontaneous breakdown of the relative U(1) phase between the two condensates.
Although SSB of the U(1) phase in a single isolated Bose-Einstein condensates is ``observed'' only theoretically, SSB in the relative U(1) phase is realistic and is directly related to experimental observations of interference.

Note that the definite relative phase $\varphi$  between the two condensates is realized, as in \rlb{BEC49} or \rlb{Theta}, by a coherent superposition of states with different divisions of particle numbers between the two subsystems.
In other words the two subsystems inevitably entangle if we demand that there is a definite relative phase.

We conjecture that the ground state obtained through the double limit in \rlb{BEC44} resembles (or coincides with) the large $L$ limit of $\ket{\Xi_\varphi^{\rm tot}}$, although the proof seems very difficult.
We also expect that states realized experimentally in weakly coupled Bose-Einstein condensates resemble  $\ket{\Xi_\varphi^{\rm tot}}$.

As an alternative approach to give a meaning of symmetry breaking in coupled Bose-Einstein condensates, a state with a fixed number of particles in a larger system (i.e., the two condensates and the environment) which is ``identical'' to  $\ket{\Xi_\theta^{\rm a}}\otimes\ket{\Xi_{\theta+\varphi}^{\rm b}}$ is constructed in \cite{ShimizuMiyadera2001}.
Here ``identical'' means that the measurement of any observable which conserves the total number of particles in the two condensates give the same results as $\ket{\Xi_\theta^{\rm a}}\otimes\ket{\Xi_{\theta+\varphi}^{\rm b}}$.
See also \cite{ShimizuMiyadera2000} for background.
It is likely that our $\ket{\Xi_\varphi^{\rm tot}}$ and their states are indistinguishable if we only measure local quantities which preserve the particle number.


\bigskip
{\small
We wish to thank Akira Shimizu and Masahito Ueda for indispensable discussions and comments which made the present work possible, and Tohru Koma and Haruki Watanabe for useful discussions on related topics.
The present work was supported by JSPS Grants-in-Aid for Scientific Research no.~16H02211.
}

\appendix
\section{Properties of the state $\ket{\Theta^{L,M}_\varphi}$}
\subsection{Expectation value of $e^{i\varphi}\had_{(x,{\rm a})}\ha_{(x,{\rm b})}$}
\label{a:ms}
Let us prove \rlb{ms2}.
Here we make use of techniques and results from \cite{Tasaki2018}.
For $\nu=\rm a,b$, let $\ho_\nu^{+}:=L^{-d}\sum_{x\in\La}\had_{(x,\nu)}$ and $\ho_\nu^{-}:=L^{-d}\sum_{x\in\La}\ha_{(x,\nu)}$.
We here abbreviate $\ket{\Theta^{L,M}_\varphi}$ as $\ket{\Theta}$.
From \rlb{Theta} and \rlb{BEC22}, we have
\eqa
\ket{\Theta}=\frac{1}{\sqrt{2M+1}}
\biggl\{\Ga\Gb&+\sum_{n=1}^Me^{in\varphi}\fnorm{(\hoa^+)^n\Ga}\fnorm{(\hob^-)^n\Gb}
\nl&+\sum_{n=1}^Me^{-in\varphi}\fnorm{(\hoa^-)^n\Ga}\fnorm{(\hob^+)^n\Gb}
\biggr\},
\lb{X}
\ena
where $\Ga$ and $\Gb$ are exact copies of the ground state $\GS$ of the Hamiltonian \rlb{BEC3B} for a single system with $N$ particles.

To evaluate the expectation value $\bra{\Theta}e^{i\varphi}\had_{(x,{\rm a})}\ha_{(x,{\rm b})}\ket{\Theta}$, we first see that 
\eqa
e^{i\varphi}\had_{(x,{\rm a})}\ha_{(x,{\rm b})}\ket{\Theta}=\frac{1}{\sqrt{2M+1}}
\biggl\{&\sum_{n=1}^{M+1}e^{in\varphi}
\fnormt{\had_{(x,{\rm a})}}{(\hoa^+)^{n-1}\Ga}
\fnormt{\ha_{(x,{\rm b})}}{(\hob^-)^{n-1}\Gb}
\nl&+\sum_{n=0}^{M-1}e^{-in\varphi}
\fnormt{\had_{(x,{\rm a})}}{(\hoa^-)^{n+1}\Ga}
\fnormt{\ha_{(x,{\rm b})}}{(\hob^+)^{n+1}\Gb}
\biggr\}.
\lb{aaX}
\ena
By using \rlb{X} and \rlb{aaX}, we get
\eqa
\bra{\Theta}e^{i\varphi}\had_{(x,{\rm a})}\ha_{(x,{\rm b})}\ket{\Theta}=&\frac{1}{2M+1}
\biggl\{\sum_{n=1}^{M+1}
\frac{
\bbkt{(\hoa^-)^n\had_{(x,{\rm a})}(\hoa^+)^{n-1}}_{\rm a}\,\,
\bbkt{(\hob^+)^n\ha_{(x,{\rm b})}(\hob^-)^{n-1}}_{\rm b}
}{
\snorm{(\hoa^+)^{n}\Ga}\snorm{(\hoa^+)^{n-1}\Ga}
\snorm{(\hob^-)^{n}\Gb}\snorm{(\hob^-)^{n-1}\Gb}
}
\nl&\hspace{1cm}+\sum_{n=0}^{M-1}
\frac{
\bbkt{(\hoa^+)^n\had_{(x,{\rm a})}(\hoa^-)^{n+1}}_{\rm a}\,\,
\bbkt{(\hob^-)^n\ha_{(x,{\rm b})}(\hob^+)^{n+1}}_{\rm b}
}{
\snorm{(\hoa^-)^{n}\Ga}\snorm{(\hoa^-)^{n+1}\Ga}
\snorm{(\hob^+)^{n}\Gb}\snorm{(\hob^+)^{n+1}\Gb}
}
\biggr\},
\lb{XaaX}
\ena
where we wrote $\sbkt{\cdots}_\nu=\bra{\Phi^\nu_{\rm GS}}\cdots\ket{\Phi^\nu_{\rm GS}}$.
Note that translation invariance implies, e.g., \par\noindent$\sbkt{(\hoa^+)^n\had_{(x,{\rm a})}(\hoa^-)^{n+1}}_{\rm a}=\sbkt{(\ho^+)^{n+1}(\ho^-)^{n+1}}$, where $\sbkt{\cdots}=\bra{\Phi_{\rm GS}}\cdots\ket{\Phi_{\rm GS}}$ and $\ho^{\pm}:=\hOL^{\pm}/L^d$ are defined for the single system.
We can thus rewrite \rlb{XaaX} as 
\eqa
\bra{\Theta}e^{i\varphi}\had_{(x,{\rm a})}&\ha_{(x,{\rm b})}\ket{\Theta}
\nl=&\frac{1}{2M+1}
\Biggl\{\sum_{n=1}^{M+1}
\frac{
\bbkt{(\ho^-)^n(\ho^+)^n}\,
\bbkt{(\ho^+)^n(\ho^-)^n}
}{
\sqrt{\bbkt{(\ho^-)^n(\ho^+)^n}\,\bbkt{(\ho^-)^{n-1}(\ho^+)^{n-1}}\,
\bbkt{(\ho^+)^n(\ho^-)^n}\,\bbkt{(\ho^+)^{n-1}(\ho^-)^{n-1}}}
}
\nl&\hspace{1.2cm}+\sum_{n=0}^{M-1}
\frac{
\bbkt{(\ho^+)^{n+1}(\ho^-)^{n+1}}\,
\bbkt{(\ho^-)^{n+1}(\ho^+)^{n+1}}
}{
\sqrt{
\bbkt{(\ho^+)^n(\ho^-)^n}\,\bbkt{(\ho^+)^{n+1}(\ho^-)^{n+1}}\,
\bbkt{(\ho^-)^n(\ho^+)^n}\,\bbkt{(\ho^-)^{n+1}(\ho^+)^{n+1}}
}
}
\Biggr\},
\nl=&\frac{2}{2M+1}
\sum_{n=1}^M\sqrt{\frac{
\bbkt{(\ho^-)^n(\ho^+)^n}\,\bbkt{(\ho^+)^n(\ho^-)^n}
}{
\bbkt{(\ho^-)^{n-1}(\ho^+)^{n-1}}\,\bbkt{(\ho^+)^{n-1}(\ho^-)^{n-1}}
}}.
\lb{XaaX2}
\ena
Following \cite{Tasaki2018}, let $\hp=(\ho^+\ho^-+\ho^-\ho^+)/2$.
Since $[\ho^-,\ho^+]=L^{-d}$, one finds that $(\ho^-)^n(\ho^+)^n=\hp^n+O(L^{-d})$ and $(\ho^+)^n(\ho^-)^n=\hp^n+O(L^{-d})$.
See Lemma~4.1 of  \cite{Tasaki2018}.
We therefore get
\eq
\bra{\Theta}e^{i\varphi}\had_{(x,{\rm a})}\ha_{(x,{\rm b})}\ket{\Theta}=
\frac{2}{2M+1}
\sum_{n=1}^M\frac{\sbkt{\hp^n}}{\sbkt{\hp^{n-1}}}+O(L^{-d}).
\lb{XaaX3}
\en
It was proved in  Lemma~4.2 of  \cite{Tasaki2018} that
\eq
\lim_{n\up\infty}\limL\frac{\sbkt{\hp^n}}{\sbkt{\hp^{n-1}}}=(\ms)^2.
\en 
This, with \rlb{XaaX3}, implies 
\eq
\lim_{M\up\infty}\limL\bra{\Theta}e^{i\varphi}\had_{(x,{\rm a})}\ha_{(x,{\rm b})}\ket{\Theta}=(\ms)^2,
\en
which is  the desired \rlb{ms2}.

\subsection{Energy expectation value}
\label{a:en}
We shall prove \rlb{LLS}.
Let $\EGS$ be the ground state energy of the Hamiltonian \rlb{BEC3B} for a single system with $N$ particles.
From \rlb{BEC23}, which shows that $\ket{\Gamma_M}$ has very low excitation energy, one readily finds that 
\eq
\limL\frac{1}{L^d}\bigl\{\bra{\Theta^{L,M}_\varphi}\hH^{\rm tot}_0\ket{\Theta^{L,M}_\varphi}-2\EGS\bigr\}=0
\lb{LLS2}
\en
for any $M$.
This is almost the desired \rlb{LLS} since it is very likely that $E^{\rm tot}_{{\rm GS},0}=2\EGS$.
But this equality cannot be proved in general and we need some work.

Define the ground state energy density (for the single system) by
\eq
\tilde{\epsilon}(\rho):=\limL\frac{\EGS}{L^d},
\en
where $L$ and $N$ always satisfy $\rho=N/L^d$.
It is standard that the limit exisits, and  $\tilde{\epsilon}(\rho)$ extends to a convex function of $\rho\in[0,1]$.
See, e.g., \cite{Ruelle,TasakiSM}.

We then note that
\eq
\limL\frac{E^{\rm tot}_{{\rm GS},0}}{L^d}=\min_{\delta}\{\tilde{\epsilon}(\rho+\delta)+\tilde{\epsilon}(\rho-\delta)\},
\en
where $E^{\rm tot}_{{\rm GS},0}$ is the ground state energy of $\hH^{\rm tot}_0=\hH_{\rm a}+\hH_{\rm b}$ with total particle number $2N$, and we again fix $\rho=N/L^d$.
But the right-hand side is equal to $2\tilde{\epsilon}(\rho)$ because of the convexity.
We have thus proved that
\eq
\limL\frac{1}{L^d}\bigl\{2\EGS-E^{\rm tot}_{{\rm GS},0}\bigr\}=0.
\lb{LLS3}
\en
Then \rlb{LLS2} and \rlb{LLS3} imply the desired \rlb{LLS}.



\begin{thebibliography}{10}

\bibitem{Dalfovo}
F. Dalfovo, S. Giorgini, L.P. Pitaevskii, and S. Stringari,
{\em Theory of Bose-Einstein condensation in trapped gases}\/,
Rev. Mod. Phys. {\bf 71}, 463--512 (1999).
\\\url{https://arxiv.org/abs/cond-mat/9806038}

\bibitem{Leggett2001}
A.J. Leggett,
{\em Bose-Einstein condensation in the alkali gases: Some fundamental concepts}\/,
Rev. Mod. Phys. {\bf 73}, 307--356 (2001).

\bibitem{BlochDalibardZwerger}
I. Bloch, J. Dalibard, and W. Zwerger,
{\em Many-body physics with ultracold gases}\/,
Rev. Mod. Phys. {\bf 80}, 885--964 (2008).
\\\url{https://arxiv.org/abs/0704.3011}

\bibitem{PitaevskiiStringari}
L. Pitaevskii and S. Stringari,
{\em Bose-Einstein condensation and superfluidity}\/ (Oxford University Press, 2016).

\bibitem{LeggettSols}
A.J. Leggett and F. Sols,
{\em On the concept of spontaneous broken gauge symmetry in condensed matter physics}\,
J. Stat. Phys. {\bf 21}, 353--364 (1991).


\bibitem{Anderson1952}
P.W. Anderson, 
{\em An Approximate Quantum Theory of the Antiferromagnetic Ground State}\/,
Phys.~Rev. {\bf 86}, 694 (1952).

\bibitem{Lhuillier}
C. Lhuillier,
{\em Frustrated Quantum Magnets}\/,
Lecture notes at ``Ecole de troisieme cycle de Suisse Romande (2002)''
\\\url{https://arxiv.org/abs/cond-mat/0502464}


\bibitem{HorschLinden}
P. Horsch, and W.~von~der~Linden, 
{\em Spin-correlations and low lying excited states of the spin-1/2 Heisenberg antiferromagnet on a square lattice}\/,
Z. Phys. {\bf B72}, 181--193 (1988).

\bibitem{KaplanHorschLinden} 
T.A. Kaplan, P.~Horsch and W.~von~der~Linden, 
{\em Order Parameter in Quantum Antiferromagnets}\/,
J.~Phys.~Soc.~Jpn.~{\bf 11}, 3894--3898 (1989).

\bibitem{KomaTasaki1993}
T. Koma and H. Tasaki,
{\em Symmetry breaking in Heisenberg antiferromagnets}\/,
Comm. Math. Phys. {\bf 158}, 191--214 (1993).
\\\url{https://projecteuclid.org/euclid.cmp/1104254136}

\bibitem{KomaTasaki1994}
T. Koma and H. Tasaki,
{\em Symmetry breaking and finite-size effects in quantum many-body systems}\/,
J. Stat. Phys. {\bf 76}, 745--803 (1994).
\\\url{https://arxiv.org/abs/cond-mat/9708132}

\bibitem{Tasaki2018}
H. Tasaki,
{\em Long-range order, ``tower'' of states, and symmetry breaking in lattice quantum systems}\/,
preprint (2018).
\\\url{https://arxiv.org/abs/1807.05847}




\bibitem{TasakiBOOK}
H. Tasaki,
{\em Physics and mathematics of quantum many-body systems}, (to be published from Springer).



\bibitem{Andrews}
M. R. Andrews, C. G. Townsend, H.-J. Miesner, D. S. Durfee,
D. M. Kurn, W. Ketterle,
{\em Observation of Interference Between Two Bose Condensates}\/,
Science {\bf 275}, 637--640 (1997).

\bibitem{Gring}
M. Gring, M. Kuhnert, T. Langen, T. Kitagawa, B. Rauer, M. Schreitl, and J. Schmiedmayer,
{\em Relaxation and prethermalization in an isolated quantum system}\/,
Science, {\bf 337}, 1318--1322  (2012).
\\\url{https://arxiv.org/abs/1112.0013}


\bibitem{Herring}
G. Herring, P. G. Kevrekidis, B. A. Malomed, R. Carretero-Gonzalez,
and D. J. Frantzeskakis, {\em Symmetry breaking in linearly coupled 
dynamical lattices}\/, Phys. Rev. E {\bf 76}, 066606 (2007).
\\\url{https://arxiv.org/abs/0704.3284}

\bibitem{Marshall}
W. Marshall,
{\em Antiferromagnetism}\/,
Proc. Roy. Soc. A {\bf 232}, 48 (1955).

\bibitem{LiebMattis1962}
E.H. Lieb and D. Mattis, 
{\em Ordering energy levels in interacting spin chains}\/, 
J. Math. Phys. {\bf 3}, 749--751 (1962).


\bibitem{KennedyLiebShastry1988a}
T. Kennedy, E.H. Lieb, and B.S. Shastry, 
{\em Existence of N\'eel order in some spin-$1/2$ Heisenberg antiferromagnets}\/,
J. Stat. Phys. {\bf 53}, 1019 (1988).

\bibitem{KennedyLiebShastry1988b}
T. Kennedy, E.H. Lieb, and B.S. Shastry, 
{\em The XY model has long-range order for all spins and all dimensions greater than one}\/,
Phys. Rev. Lett. {\bf 61}, 2582 (1988).

\bibitem{KuboKishi1988} 
K. Kubo and T. Kishi, 
{\em Existence of long-range order in the XXZ model}\/,
Phys. Rev. Lett. {\bf 61}, 2585 (1988).

\bibitem{DysonLiebSimon1978}
F.J. Dyson, E.H. Lieb, B. Simon,
{\em Phase transitions in quantum spin systems with isotropic and nonisotropic interactions}\/, 
J. Stat. Phys. {\bf 18}, 335--382 (1978).

\bibitem{NevesPerez1986} 
E.J. Neves and  J.F. Perez, 
{\em Long range order in the ground state of two-dimensional antiferromagnets}\/,
Phys. Lett. {\bf 114}A, 331-333 (1986).



\bibitem{RingSchuck}
 P. Ring and P. Schuck,
 {\em The Nuclear Many-Body Problem}\/, (Springer, 1980).


\bibitem{Mattis79}
D. Mattis,
{\em Ground-state symmetry in XY model of magnetism}\/,
Phys. Rev. Lett. {\bf 42}, 1503 (1979).

\bibitem{Nishimori81}
H. Nishimori,
{\em Spin Quantum Number in the Ground State of the Mattis-Heisenberg Model}\/,
J. Stat. Phys. {\bf 26}, 839--845 (1981).

\bibitem{FetterWalecka}
A.L. Fetter and J.D. Walecka,
{\em Quantum Theory of Many-Particle Systems}\/, (Dover, 2003).


\bibitem{Javanainen}
J. Javanainen and S.M. Yoo,
{\em Quantum Phase of a Bose-Einstein Condensate with an Arbitrary Number of Atoms}\/, 
Phys. Rev. Lett. {\bf 76}, 161 (1996).

\bibitem{ShimizuMiyadera2001}
A. Shimizu and T. Miyadera,
{\em Charge superselection rule does not rule out pure states of subsystems to be coherent superpositions of states with different charges}\/, preprint (2001).
\\\url{https://arxiv.org/abs/cond-mat/0102429}

\bibitem{ShimizuMiyadera2000}
A. Shimizu and T. Miyadera,
{\em Robustness of Wave Functions of Interacting Many Bosons in a Leaky Box}\/,
Phys. Rev. Lett. {\bf 85}, 688--691 (2000). Errata: Phys. Rev. Lett. {\bf 86}, 4422 (2001).



\bibitem{Ruelle}
D. Ruelle,
{\em Statistical Mechanics: Rigorous Results}\/,
(World Scientific, 1999).

\bibitem{TasakiSM}
H. Tasaki, {\em Statistical Mechanics}\/ (in Japanese), (Baifukan, 2008).
The English version by H. Tasaki and G. Paquette is in preparation.




\end{thebibliography}
\end{document}